\documentclass[aip, pof]{revtex4-2}

\usepackage{graphicx}
\DeclareGraphicsExtensions{.pdf,.png,.jpg}
\usepackage{dcolumn}
\usepackage{bm}
\usepackage[utf8]{inputenc}
\usepackage[T1]{fontenc}
\usepackage{amssymb}
\usepackage{mathptmx}
\usepackage{color}
\usepackage[table]{xcolor} 
\usepackage{longtable}
\usepackage{array}           
\usepackage{booktabs}        
\newcolumntype{P}[1]{>{\centering\arraybackslash}p{#1}}
\newcommand{\corr}[1]{{\color{black} #1}}
\usepackage[hidelinks]{hyperref}

\begin{document}


\title[Aggregates in fluidized beds: the effect of bonding angles on fluidization]{Aggregates in fluidized beds: the effect of bonding angles on fluidization\\	
This paper appeared in AIP Advances, 15, 105016, 2025, and was published by AIP Publishing under the CC BY 4.0 license. It can be found on https://doi.org/10.1063/5.0293132} 



\author{Vin\'icius P. S. Oliveira}
\affiliation{Faculdade de Engenharia Mec\^anica, Universidade Estadual de Campinas (UNICAMP),\\
	Rua Mendeleyev, 200, Campinas, SP, Brazil
}

\author{Danilo S. Borges}
 \email{proffisicadaniloborges@gmail.com}
\thanks{Corresponding author}
\affiliation{Faculty of Physics, University of Duisburg-Essen,
	Lotharstrasse 1, 47057 Duisburg, Germany
}

\author{Erick M. Franklin}%
\affiliation{Faculdade de Engenharia Mec\^anica, Universidade Estadual de Campinas (UNICAMP),\\
	Rua Mendeleyev, 200, Campinas, SP, Brazil
}


\date{\today}

\begin{abstract}
	
Fluidized beds consist of solid particles suspended in a tube by an ascending fluid. In liquids, it is not rare that particles adhere to each other, decreasing the solid-liquid contact area and the ratio between the tube and grain diameters, deteriorating fluidization. We inquire into this problem by carrying out experiments with trios of spheres fluidized by water flows, the spheres being glued in predefined angles. In our tests, we used a 25.4-mm-ID (internal diameter) tube and 5.95-mm-diameter spheres, and we varied the angle of trios within 60$^{\circ}$ and 180$^\circ$ and water velocities within 0.027 and 0.210 m/s. Due to the small ratio between the diameters of the tube and spheres (approximately 4.3), the bed is prone to the formation of plugs and clogs. Our experiments show that elutriation, fluidization with plugs, glass transitions (amorphous static structures), packed beds, clogging, and a transitional clogged-plug regime can appear in the bed, depending on the bonding angles and water velocities. We report the relations between the bed height, bonding angles, and flow velocity, and show that they correlate with the granular temperature. We also show that an angle of 90$^\circ$ maximizes fluidization for a given fluid velocity, and we propose a regime map that organizes the different patterns based on the bonding angle and flow velocity. The proposed map can serve as a guide for selecting the fluid velocities in order to keep the bed fluidized at all times, helping in the design and operation of fluidized beds.

\end{abstract}

\pacs{}

\maketitle 

\section{INTRODUCTION}
\label{sec:intro}

A fluidized bed is basically a suspension of granular material by an ascending fluid inside a tube, in which the balance of forces assures that the weight of the granular column is, in average, in equilibrium with the forces caused by the fluid flow \cite{Davidson, Kunii}. Due to fluctuations caused by the fluid flow, the grains oscillate and move while the bed is fluidized, with grain-grain and grain-wall collisions taking place in many cases (usually when the flow velocity is strong enough) \cite{Davidson, guazzelli_book}. Because of the relative high area between the solid and the fluid, in addition to the oscillations and collisions that take place, fluidized beds have high rates of mass and heat transfers between solids and fluids in a compact volume, being frequently employed in industrial activities \cite{Kunii}. Some examples are the drying of seeds \cite{Yang}, the coating of beads \cite{Yang, Saleh}, and the combustion and gasification of coal and biomass \cite{Yang, Basu}. 

For an optimal operation, the particles typically have sizes within 0.1 mm and 1 cm, depending on the fluid \cite{Geldart}: on the one hand, they must be small enough to increase the contact area between the fluid and the solid; on the other hand, they must be large enough to have considerable inertia with respect to the fluid, ensuring that they will not be elutriated (conveyed further downstream by the fluid) and will oscillate and collide with each other \cite{Kunii, Yang}. Therefore, it is common practice to grind the solid into smaller pieces that will form the bed to be fluidized.

However, it is not rare that solid particles adhere to each other and remain bonded. In the case of solid-liquid fluidized beds (SLFBs), there are at least two mechanisms that have been reported. One important mechanism occurs in micro SLFBs (tube diameter $D$ $\lessapprox$ 3 mm), where adhesion forces can reach values of the
same order of hydrodynamic and gravitational forces \cite{Nascimento}. Micro SLFBs are employed in the capture of $CO_2$ \cite{Fang, Shen}, gasification \cite{Zeng, Zhang2, Cortazar}, encapsulation \cite{Schreiber, Rodriguez}, pyrolysis \cite{Jia, Gao, Mao, Yu}, and wastewater treatment \cite{Kuyukina, Kwak}, just to cite a few examples, and the clusters of particles and clogs that appear due to adhesion deteriorate the desired processes. The other mechanism consists in the formation of a film of biological material over each particle, which then adhere to each other and form groups of two or more particles that remain bonded \cite{Dempsey}. One typical example is the fluidized-bed bioreactors, employed for  biological treatment of domestic wastewater in places not covered by extensive water treatment \cite{Dempsey,Nelson}, and, as in the case of micro SLFBs, the formation of clusters and clogs deteriorates the desired processes.

Besides the reduction in oscillations, collisions, and contact area, the bonding of particles generates aggregates that, as a whole, can be seen as large grains. If tube diameter $D$ is not much larger than typical length of loose particles (the diameter $d$, in case of spheres), then the resulting confinement is strong and can lead to clogs, sedimentation and jamming \cite{Cunez4}. This is the case of narrow and very-narrow beds, for which we consider 10 $\leq$ $D/d$ $\leq$ 100 and $D/d$ $<$ 10, respectively \cite{Cunez3} (there is not a real consensus in this terminology). These beds are susceptible to instabilities and patterns that are different from those in usual beds \cite{Didwania, Duru, Ghatage}, and, in particular, C\'u\~nez and Franklin \cite{Cunez, Cunez3} showed that structures in the form of granular plugs, crystallization, and jamming can occur in very-narrow beds. In those works, crystallization has been defined as the absence of motion at the bed (macroscopic) scale, and jamming as the absence of motion at both the bed and grain (microscopic) scales. In addition, the formation of small bonded structures can reach a length equal to the tube diameter, deteriorating significantly fluidization. Very-narrow fluidized beds are typically used in the examples cited for micro SLFBs, being particularly important in the biological treatment of domestic wastewater. For that, SLFB bioreactors are used \cite{Dempsey, Nelson}, which contains particles that are already relatively large with respect to the tube diameter and that adhere to each other (increasing confinement even more).

There are few investigations of usual and narrow beds with bonded particles \cite{Mikami, Kuwagi, Zhou4, He, Bahramian}, but those of very-narrow SLFBs are even scarcer, the only reported work being, to the best of our knowledge, C\'u\~nez et al. \cite{Cunez4}. In that paper, the authors carried out an experimental and numerical investigation of beds consisting of either pairs of spheres (duos) or sets of three spheres (trios) bonded together. For the trios, the same configuration with all particles touching each other (60$^\circ$ internal angle, see Fig. \ref{fig:1}b) was used. The authors used aluminum spheres with $d$ $=$ 4.8 mm in a tube with $D$ $=$ 25.4 mm, so that $D/d$ = 5.3 (ratio still lower if we consider the typical length of the aggregates), and the fluid was water. For the experiments, they used image processing and for the simulations CFD-DEM (computation fluid dynamics - discrete element methods).  C\'u\~nez et al. \cite{Cunez4} found that granular plugs appear, with an average length $\lambda$ that decreases with the water velocity, just as in the case of loose grains, but that the aggregates remain bounded in some regions, different from the loose grains (which travel all the tube, passing from plug to plug). In the case of trios, they found that they can group and form clogs at some distance from the tube entrance, and that these clogs can jam (no motion of grains even in the microscopic scale) after some time has elapsed. When this happens, the objective of a fluidized bed is greatly compromised.

Although C\'u\~nez et al. \cite{Cunez4} measured the motion of bonded particles at both macroscopic and microscopic scales, showing the bed patterns, typical trajectories of grains, and granular temperature, they did not investigated the effect of different structures of trios on fluidization and defluidization. In this paper, we inquire further into the effects of bonded particles on very-narrow SLFBs, investigating the bed response to different structures of aggregates. \corr{More specifically, we are neither investigating the breaking up nor the formation of aggregates, but how the different geometries of trios affect fluidization.} For that, we carried out experiments with trios of spheres fluidized by water flows, in which the spheres were glued in a predefined angle, always forming planar structures. We used a 25.4-mm-ID tube, the spheres had a mean diameter of 5.95 mm ($D/d$ $=$ 4.3), and we varied the angle of trios within 60$^\circ$ and 180$^\circ$ and the water velocities within 0.027 and 0.210 m/s. \corr{The determination and measurements of both bed patterns and motion of particles (mainly in terms of granular temperature) were made based on images acquired with a digital camera and image processing. With that, we found, for the first time, the regimes in which the bed is fully fluidized and those in which it defluidizes, deteriorating fluidization. We show} that elutriation (conveying of particles), fluidization with plugs, glass transitions (amorphous static structures), packed beds, clogging, and a transitional clogged-plug regime can appear in the bed, depending on the bonding angles and water velocities, and we report the relations between the bed height, angle of particles, and flow velocity. We also show that an angle of 90$^\circ$ maximizes fluidization (among trios) for a given fluid velocity, and we propose a regime map that organizes the different patterns based on the bonding angle and flow velocity. Therefore, different from C\'u\~nez et al. \cite{Cunez4}, we show that the bonding angle influences significantly the behavior of the bed, and find the different regimes as functions of such angle and water velocities. Our findings bring new insights into the spontaneous defluidization of beds in which cohesion and adhesion take place, such as those used in chemical processes and water treatment.

In the following, Sec. \ref{sec:exp_setup} presents the experimental setup, Sec.  \ref{sec:results} shows the results, and Sec. \ref{sec:conclusions} concludes the paper.

\section{EXPERIMENTAL SETUP}
\label{sec:exp_setup}

\begin{figure}[ht]
	\centering
	\includegraphics[width=0.7\columnwidth]{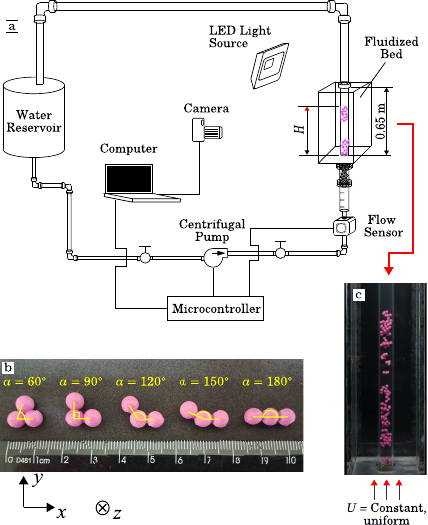}
	\caption{(a) Layout of the experimental setup. (b) Angles $\alpha$ used in the trios. (c) snapshot of the test section (with the visualization box), showing one example of fluidized bed.}
	\label{fig:1}
\end{figure} 

The experimental setup consisted of a water tank, a centrifugal pump, a flowmeter, a flow homogenizer, a 25.4-mm-ID vertical tube 1.2 m long, a return line, and some valves and fittings, as shown in Fig. \ref{fig:1}a. The vertical tube was made of transparent material (polymethyl methacrylate - PMMA), and its first 0.65 m corresponds to the test section. It was aligned vertically within $\pm 3^{\circ}$, and the test section was enclosed in a rectangular box filled with water (visualization box) to mitigate optical distortions. The flow homogenizer, placed upstream the test section, consisted of a 150-mm-long tube filled with a packed bed of $d$ = 6 mm spheres. The water flow rate was controlled by varying the pump rotation via a frequency inverter that was connected to a computer, in its turn connected to the flowmeter. A photograph of the test section is shown in Fig. \ref{fig:1}c, and a photograph of the experimental setup is available in the supplementary material.

We used tap water within 25$^{\circ}$C $\pm$ 3$^{\circ}$C, with density $\rho_f$ $\approx$ 1000 kg/m$^3$ and dynamic viscosity $\mu_f$ $\approx$ 10$^{-3}$ Pa.s, and 0.2 g polymer-covered spheres with diameter $d$ = 5.95 mm $\pm$ 0.01 mm and $\rho_p$ = 1768 kg/m$^3$. Therefore, by considering single (loose) spheres, $D/d$ $=$ 4.3, corresponding to a very-narrow SLFB. Still for single spheres, the number of Stokes based on the terminal velocity $v_t$ was $St_t \,=\, v_t d \rho_p / (9\mu_f)$ $=$ 404, and the number of Reynolds based on $v_t$ was $Re_t \,=\, \rho_f  v_t d / \mu_f$ = 2056, indicating significant inertia of the grains with respect to the fluid. Sets of three spheres (trios) were glued together by using molds that positioned grains with a given angle between contacts (on a plane), forming the trios shown in Fig. \ref{fig:1}b. We used a small amount of epoxy glue for bonding the spheres (just enough to bond spheres together at the contact point between consecutive spheres, not covering their entire surface), which increased the total mass of trios by only $3\%$, and the angle between contacts were $60^\circ$, $90^\circ$, $120^\circ$, $150^\circ$, and $180^\circ$. With this procedure, the trios were stable structures that did not break up into duos or loose grains, and the angles were kept constant throughout tests (the trios were rigid structures, the main objective of this work being to understand how the different geometries of trios affect fluidization). With that, we produced five different types of trios, sharing the same mass, surface area, and volume, but with different components of the moment of inertia. The values of the horizontal and vertical lengths, $l_x$ and $l_y$, respectively, of the $xx$, $yy$ and $zz$ components of the moment of inertia, $I_{xx}$, $I_{yy}$, and $I_{zz}$, respectively, of the aspect ratio $\lambda$ $=$ $l_y/l_x$, of the local packing fraction $\phi_{l}$ computed in a bounding box involving the particles, and minimum projected area $min(A_p)$, are shown in Tab. \ref{table:table1}.

\begin{table}[ht]
	\caption{For each angle $\alpha$, values of the horizontal and vertical lengths, $l_x$ and $l_y$, respectively, of the $xx$, $yy$ and $zz$ components of the moment of inertia, $I_{xx}$, $I_{yy}$, and $I_{zz}$, respectively, of the aspect ratio $\lambda$ $=$ $l_y/l_x$, of the local packing fraction $\phi_{l}$ computed in a bounding box involving the particles, and minimum projected area $min(A_p)$.}
	\label{table:table1}
	\centering
	\begin{tabular}{c c c c c c c c c}  
		\hline\hline
		$\alpha$ &  $l_x$  & $l_y$ & $\lambda$ & $\phi_{l}$ & $I_{xx}$  & $I_{yy}$ & $I_{zz}$ & $min(A_p)/(\pi d^2/4)$\\
		
		($^{\circ}$) &  (mm)  & (mm) & $\cdots$ &  $\cdots$ & (g$\cdot$ mm$^2$) & (g$\cdot$ mm$^2$) & (g$\cdot$ mm$^2$) & $\cdots$\\
		\hline 
		60   & 11.90 & 11.10  & 0.93 & 0.63 & 3.01 & 3.01 & 3.89 & 1.80  \\
		90   & 11.90 & 11.90  & 1.00 & 0.59 & 2.71 & 3.89 & 4.48 & 1.66 \\
		120  & 14.88 & 11.10  & 0.75 & 0.50 & 2.42 & 4.78 & 5.07 & 1.47 \\
		150  & 17.05 & 8.93  & 0.52 & 0.55 & 2.20 & 5.43 & 5.51 & 1.25 \\
		180  & 11.90 & 5.95  & 0.33 & 0.78 & 2.12 & 5.66 & 5.66 & 1.00 \\
		\hline
		\hline 
	\end{tabular}
\end{table}

\begin{table}[h]
\centering
\caption{Mean bed height $H_{if}$ and the corresponding standard deviation $\sigma_{H_{if}}$ at incipient fluidization, initial average packing fraction $\phi_0$ and the corresponding standard deviation $\sigma_{\phi_0}$, water velocity $U_{if}$  at incipient fluidization, and the settling velocity of spheres $v_{s,i}$, for different bonding angles $\alpha$.}
\label{tab:incipient}
\color{black} 
\begin{tabular}{m{1cm} m{1.5cm} m{1.7cm} m{1.4cm} m{1.4cm} m{1.8cm} m{1.7cm}}
	\hline
	\hline
	$\alpha$ ($^{\circ}$) & $H_{if}$ (m) & $\sigma_{H_{if}}$ & $\phi_0$ & $\sigma_{\phi_0}$ & $U_{if}$ (m/s) & $v_{s,i}$ (m/s)\\
	\hline 
	60 & 0.066 & 0.005 & 0.495 & 0.047 & 0.093 & 0.067\\ 
	90 & 0.096 & 0.062 & 0.361 & 0.033 & 0.110 & 0.127\\ 
	120 & 0.081 & 0.002 & 0.412 & 0.051 & 0.099 & 0.099\\ 
	150 & 0.085 & 0.011 & 0.386 & 0.026 & 0.099 & 0.108\\ 
	180 & 0.089 & 0.040 & 0.376 & 0.020 & 0.110 & 0.115\\ 
	\hline
	\hline
\end{tabular}
\end{table}

In each test run the bed consisted of 150 trios of one single type, for which we varied the cross-sectional average velocities $U$ (also called superficial velocities) within 0.027 and 0.210 m/s. For each bed type, Tab. \ref{tab:incipient} presents the average particle fraction $\phi_0$ of the settled bed, the heights $H_{if}$ and water velocities $U_{if}$ at the inception of fluidization, and the settling velocity of spheres $v_{s,i}$. As in previous works \cite{Cunez3, Cunez4}, values of $\phi_0$, $H_{if}$ and $U_{if}$ were determined experimentally by using image processing, and those of $v_{s,i}$ by using the Richardson--Zaki correlation (for reference only), $v_{s,i} = v_t \left( 1-\phi_0 \right) ^{2.4}$. We note that the incipient condition is determined based on the beginning of particle motion (small oscillations), and that the use of higher $D/d$ would imply different values of $\phi_0$ (but this is beyond the scope of this paper). Prior to each test, grains were let to settle by free fall in the test section, and just after that the tests began by increasing the water flow until reaching a pre-defined water velocity (that remained constant during all the test). Each test had a total duration of 300 s, and the same tested condition was repeated five times.

We placed a camera aligned in the horizontal direction (perpendicularly to the test section), with a lateral view of the bed (as shown in Fig. \ref{fig:1}a). The camera was of complementary metal-oxide-semiconductor (CMOS) type, with a maximum resolution of 1920 px $\times$ 1080 px when operating at 60 Hz, and we mounted a lens of $60$ mm focal distance and F2.8 maximum aperture on the camera. The same computer used to control the pump rotation was also used to control the camera.  We set the camera frequency to 30 Hz and the region of interest (ROI) to 100 px $\times$ 1330 px, for a field of view of 25.4 mm $\times$ 337.8 mm, so that 1 px corresponds to approximately 0.25 mm in the images. Finally, we used lamps of light-emitting diode (LED) branched to a continuous-current source to avoid beating between the light and the camera frequencies, and we placed a black background to highlight the contrast between the particles (that were pink) and the background. The acquired images were then processed with numerical codes developed by ourselves (the numerical scripts and some figures used in the present work are available in an open repository \cite{Supplemental4}).

Finally, we note that we are investigating the effect of bonding angles on fluidization for the first time. Therefore, we decided to reduce the number of parameters, limiting the number of grains to three and the angles to those in a plane, so that we can better understand the problem before testing more complex scenarios in a future work.

\section{RESULTS}
\label{sec:results}

From the acquired images, we assessed both the macroscopic (bed scales) and the microscopic (grain scale) quantities. We observed that, depending on the angles of particles and the flow velocity, different patterns arise, and that those patterns correlate with the level of fluctuation at the grain scale.

\subsection{Macroscopic scale}
\label{subsec:macroscopic}

\begin{figure}[ht]
	\centering
	\includegraphics[width=0.9\columnwidth]{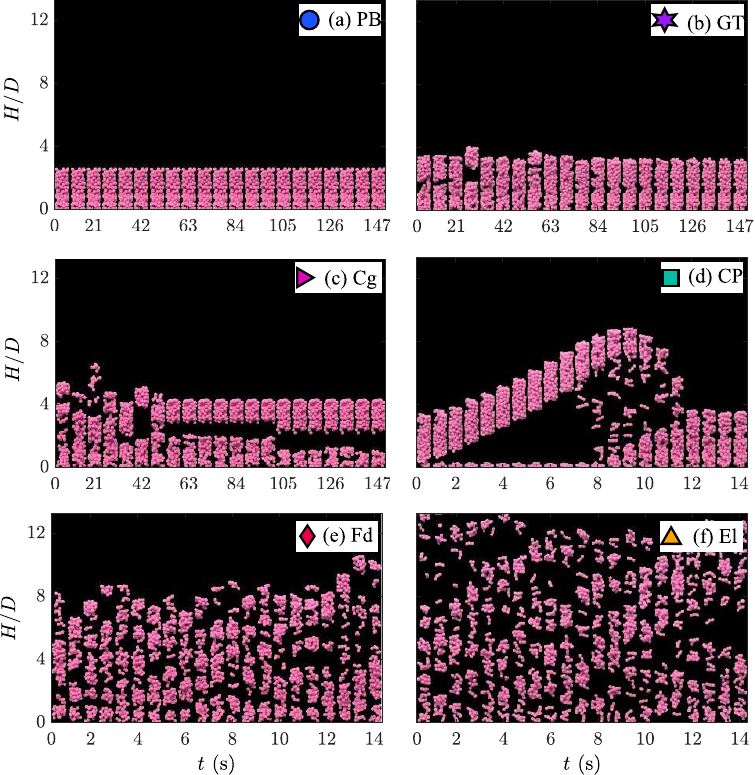}
	\caption{Snapshots of the bed placed side by side: (a) packed bed ($\alpha$ $=$ 120$^\circ$,  $U$ $=$ 0.055 m/s); (b) glass transition ($\alpha$ $=$ 150$^\circ$,  $U$ $=$ 0.099 m/s); (c) clogging ($\alpha$ $=$ 120$^\circ$,  $U$ $=$ 0.121 m/s); (d) clogged-plug ($\alpha$ $=$ 90$^\circ$,  $U$ $=$ 0.110 m/s); (e) fluidized ($\alpha$ $=$ 120$^\circ$,  $U$ $=$ 0.137 m/s); (f) elutriation ($\alpha$ $=$ 150$^\circ$,  $U$ $=$ 0,192 m/s). The time interval between each snapshot is 7 s in panels a-c and 0.7 s in panels d-f. (multimedia available online)}
	\label{fig:snapshots}
\end{figure}

In general, we observed the formation of granular plugs that move upward, the sedimentation of grains under some conditions (crystallization or glass transition), and the formation of clogged regions, as already reported in previous works \cite{Cunez, Cunez3, Cunez4, Oliveira}. But now we inquire also into other observed patterns, such as the packed bed, elutriation, and an intermediate state that we name clogged plug. In order to better organize the results, we call in the following those regimes packed bed, clogging, fluidized, clogged-plug, glass transition, and elutriation, and we describe each one of them next.  Figure \ref{fig:snapshots} shows snapshots of the bed placed side by side, illustrating the different observed patterns (multimedia available online). A dimensional version of this figure is available in the supplementary material.

The packed bed (PB) regime (Fig. \ref{fig:snapshots}a) corresponds to a bed that is not fluidized, obtained by increasing the water velocity from zero until reaching the limit for fluidization. Therefore, it happens when the water velocity is insufficient for balancing the particles' weight and disrupting the force chains of the packed trios, suspending the particles. It takes place, thus, for superficial velocities $U$ lower than that necessary for incipient fluidization $U_{if}$, the latter depending on the bonding angle $\alpha$ (Tab. \ref{tab:incipient}). In addition, we observe that for each $\alpha$, there exists an incipient bed height $H_{if}$, which varies in a non-monotonic way, as shown in Tab. \ref{tab:incipient}. In other words, there is a value of $\alpha$ that maximizes $H_{if}$.

The glass transition (GT, also referred as crystallization in previous papers, \cite{Cunez3, Oliveira}), shown in Fig. \ref{fig:snapshots}b, occurs when the bed defluidizes spontaneously at velocities above those for incipient or minimum fluidization \cite{Goldman, Cunez3, Oliveira}. It is characterized by the settling of grains, that then form a static structure that presents no motion at the bed scale (macroscopic), but have small fluctuations at the scale of grains (microscopic). If the water velocity is increased once glass transition has taken place, the bed can jam: even the microscopic oscillations become negligible \cite{Goldman, Cunez3}. In our experiments, the GT regime occurs spontaneously with $U$ held constant throughout the test, and was observed only within $150^{\circ}$ $\le$ $\alpha$ $\le$ $180^{\circ}$.

The fluidized regime (Fd), shown in Fig. \ref{fig:snapshots}e, occurs for velocities above that for incipient fluidization, $U_{if}$, and has a finite duration in cases when the bed transitions to GT, clogging (Gg, described next), or clogged-plug (CP, described next). In those cases, the limited fluidization occurs for velocities within  $U_{if}$ $\leq$ $U$ $<$ $U_{cont}$, where $U_{cont}$ is the minimum velocity for keeping fluidization for an indeterminate duration, and we classify them as GT, Cg, and CP (described next) instead of Fd. Above $U_{cont}$ and below $U_{el}$, where $U_{el}$ is the minimum velocity for elutriation (descibed next), the bed remains fluidized at all times, which we then classify as Fd. In the Fd regime, the fluid forces balance, in average, the weight of the granular column and its interactions with the tube wall. In the case of very-narrow beds, this regime is characterized by granular plugs that move upwards \cite{Cunez, Cunez3, Cunez4, Oliveira}, with the grains migrating from one plug to another. When $U_{cont}$ $\leq$ $U$ $<$ $U_{el}$, the Fd regime was observed for all values of $\alpha$.

The elutriation regime (El), shown in Fig. \ref{fig:snapshots}f, corresponds to an hydraulic conveying of particles, i.e., they are entrained further downstream by the water flow. It takes place for water velocities higher than $U_{el}$, for which the forces caused by the fluid become higher than those of the granular column and its interactions with the tube wall. Given the range of $U$ used in our experiments, the El regime was observed only for $120^{\circ}$ $\leq$ $\alpha$ $\leq$ $150^{\circ}$, but they are expected to occur for all values of $\alpha$ if values of $U$ are high enough (in other words, our results show that the values of $U_{el}$ are smaller for $120^{\circ}$ $\leq$ $\alpha$ $\leq$ $150^{\circ}$).

The clogging regime (Cg), shown in Fig. \ref{fig:snapshots}c, corresponds to the spontaneous formation of localized structures that obstruct the flow, not necessarily taking place close to the tube entrance. Traditionally, clogging has been studied in bottleneck flows, with fewer studies dealing with clogging in narrow vertical pipes \cite{Janda, Alvarez, Lopez}, and ever fewer in the case of bonded particles \cite {Cunez4}. In the present case, the clog has an arch-like structure that hinders the motion of its grains while pushed in the downstream direction by the fluid flow (since forces are redirected from the vertical to the horizontal direction via solid-solid contacts). Therefore, in order to break the clog and restore the motion of its grains, it is necessary to decrease the fluid flow below $U_{if}$. In our experiments, clogs appeared only for $120^{\circ}$ $\leq$ $\alpha$ $\leq$ $150^{\circ}$, though C\'u\~nez et al. \cite{Cunez4} reported clogging for $\alpha$ $=$ 60$^{\circ}$ in beds consisting of 100-200 trios of aluminum beads. This suggests that the probability of clogging depends on the properties of particles, their number, and their aspect ratio, so that for a given value of $\alpha$ there would exist a critical number of trios for clogging. However, we do not investigate this in this paper.

Finally, in the clogged-plug regime (CP), shown in Fig. \ref{fig:snapshots}d, there is the formation of a clog that slides in the flow direction (different from Cg, CP involves motion of the entire structure). The CP structure is similar to that of the Gg regime, with an arch-like structure being formed at some point after the bed has fluidized. However, in the CP regime the redirection of forces seems to be somewhat weaker than that for the Cg regime, so that the friction force at the contact between the peripheral particles and the tube wall is overcome by the forces due to the fluid flow. Because the CP structure moves upward, at some point its lower particles are released (settling by free fall) and the CP eventually disappears (as can be observed from $t$ $=$ 7 s in Fig. \ref{fig:snapshots}d). We note that, apart from the break up stage, grains in the CP structure maintain their relative position with respect to one another. The CP structure is robust, migrating with a non-constant velocity (depicted as a green line following a nonlinear path in Fig. 2 of the supplementary material) and covering a significant distance (at least two times its initial length). We also observe a slight compaction of the CP structure during its upward motion, and that the CP regime occurs for all values of $\alpha$, but more intensely for lower bonding angles (as shown next).

\begin{figure}[ht]
	\centering
	\includegraphics[width=0.7\columnwidth]{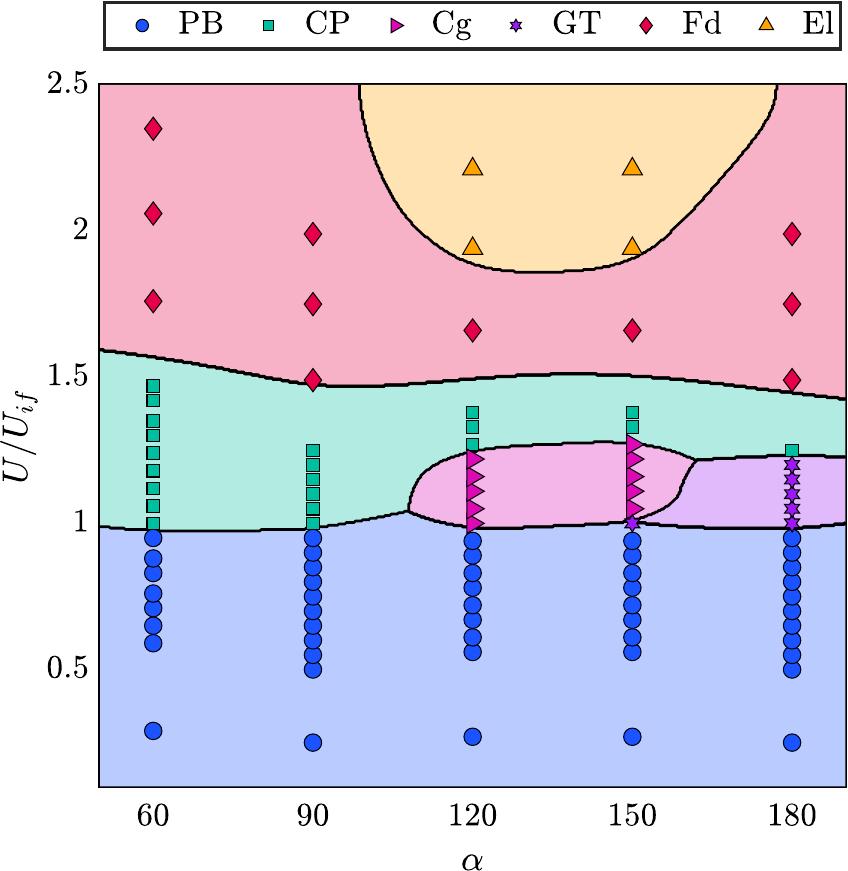}
	\caption{Regime map in the $\alpha$ -- $U/U_{if}$ space. The symbols are listed in the figure key, and the curves separating the patterns were drawn using SVM (support vector machine).}
	\label{fig:regime_map}
\end{figure} 

The observation of these six regimes raises the question of how transitions from one regime to another happen, and what are the important parameters controlling them. The flow velocity is one important parameter controlling the PB (velocities below $U_{if}$) and El (velocities above $U_{el}$) regimes, but, in their turn, $U_{if}$  and $U_{el}$ depend on the bonding angle $\alpha$. Therefore, $\alpha$ is another important parameter. For intermediate velocities ($U_{if}$ $\leq$ $U$ $<$ $U_{el}$), the Fd, GT, Cg, and CP regimes can appear, and some of these regimes appear only for given values of $\alpha$. We choose, thus, $U/U_{if}$ and $\alpha$ as control parameters for this problem, and build an ad hoc regime map in the $\alpha$ -- $U/U_{if}$ space, shown in Fig. \ref{fig:regime_map}.

For the smaller velocities, the packed bed regime persists for all types of trios as long as $U$ $\leq$ $U_{if}$, but the incipient velocity $U_{if}$ depends on the bonding angle $\alpha$, as shown in Tab \ref{tab:incipient}. Variations of $U_{if}$ with the bonding angle are, however, rather small, and the ranges of $U/U_{if}$ for the PB regime are roughly constant across different values of $\alpha$. For $U$ $>$ $U_{el}$, we expect that the El regime occurs for all bonding angles. However, given the range of $U$ used in our experiments, the El regime was observed only for $120^{\circ}$ $\leq$ $\alpha$ $\leq$ $150^{\circ}$. This basically means that the values of $U_{el}$ are smaller for $120^{\circ}$ $\leq$ $\alpha$ $\leq$ $150^{\circ}$, and we did not reach $U$ $>$ $U_{el}$ for $\alpha$ $\leq$ $90^{\circ}$ or $\alpha$ $=$ $180^{\circ}$.

For $U$ $>$ $U_{if}$, we observe that the CP regime appears for all values of $\alpha$, but it is more intense as $\alpha$ decreases. For $\alpha$ $\leq$ $90^{\circ}$, CP takes place just after the PB regime (by increasing $U$), persisting until $U/U_{if}$ $\approx$ 1.5 and $\approx$ 1.3 for $\alpha$ $=$ $60^{\circ}$ and $\alpha$ $=$ $90^{\circ}$, respectively. For $\alpha$ $\geq$ $120^{\circ}$, another regime appears as $U$ becomes larger than $U_{if}$, after which the CP regime appears, with decreasing ranges of $U/U_{if}$ as $\alpha$ increases. The Fd regime appears when $U$ $<$ $U_{el}$ and $U$ is greater than the range for the CP regime, for all values of $\alpha$. The minimum values of $U/U_{if}$ are lower for $\alpha$ $=$ $90^{\circ}$ and $\alpha$ $=$ $180^{\circ}$, and the higher values are lower for $120^{\circ}$ $\leq$ $\alpha$ $\leq$ $150^{\circ}$ (due to lower values of $U_{el}$, as discussed in the El case). The GT regime appears as $U$ becomes larger than $U_{if}$, after which the Cg regime appears for $\alpha$ $=$ $150^{\circ}$ and the CP regime for $\alpha$ $=$ $180^{\circ}$, the GT regime being more intense in this latter case (the range of $U/U_{if}$ is much larger for $\alpha$ $=$ $180^{\circ}$). Finally, the Cg regime appears between the PB and CP regimes for $\alpha$ $=$ $150^{\circ}$, and between the GT and CP regimes for $\alpha$ $=$ $180^{\circ}$. 

\begin{figure}[ht]
	\centering
	\includegraphics[width=0.7\columnwidth]{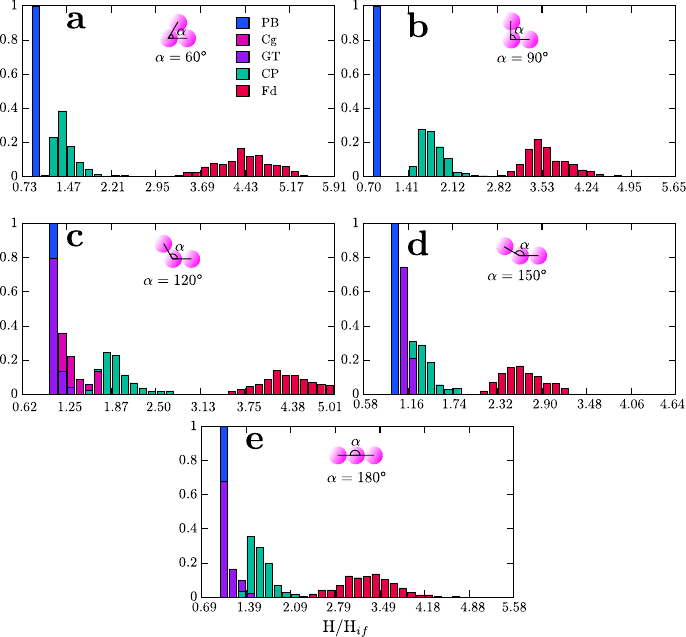}
	\caption{Histograms showing the distributions of bed heights $H/H_{if}$ measured for the different types of trios (bonding angles $\alpha$ are indicated in the key of figures).}
	\label{fig:bedheight_angles}
\end{figure}

By considering that the main goal of a fluidized bed is to maintain the grains suspended without carrying them further downstream, and to keep them as loose as possible, we observe from Fig. \ref{fig:regime_map} that the best conditions occur when $\alpha$ $=$ 90$^\circ$. In order to investigate further the bed structure, we analyze next the bed heights $H$ reached for the different angles tested and regimes observed. Figure \ref{fig:bedheight_angles} shows the histograms of the distributions of bed heights $H$ for the different types of trios, normalized by the height at the incipient fluidization $H_{if}$ (histograms of $H$ in dimensional form are available in the supplementary material). We observe that the highest mean values are obtained in the Fd regime, being higher for $\alpha$ $=$ 60$^\circ$ and 120$^\circ$, which implies lower average packing fractions $\phi$ for these values of $\alpha$. However, distributions are significantly spread, so that values for $\alpha$ $=$ 90$^\circ$ and 180$^\circ$, also reach high values of $H$ (indicating also low average packing fractions and ``good'' fluidization). Well below the Fd regime, we find the CP regime, with heights corresponding to 1/3 and 1/2 of those of Fd (meaning packing fractions 2-3 times higher). Although there is macroscopic motion in the CP regime, the high compactness of its structure implies much worse heat and mass transfers when compared with the Fd regime. The PB, of course, has the lowest heights (and higher packing fractions), and the GT and Cg regimes follow closely, with also small values. Therefore, as expected, the PB, GT, and Cg regimes are to be avoided if the goal is suspend particles to achieve high rates of heat and mass transfers between the solid and the fluid.

\begin{figure}[ht]
	\centering
	\includegraphics[width=0.6\columnwidth]{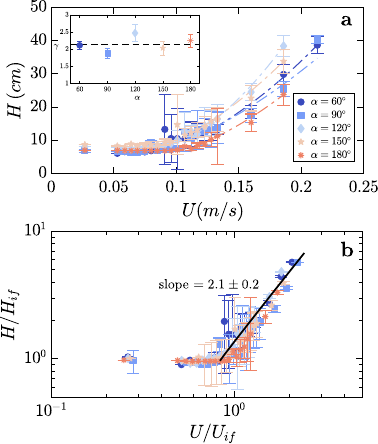}
	\caption{Bed height $H$ versus superficial velocity $U$ in (a) dimensional form, and (b) dimensionless form (using $H_{if}$ and $U_{if}$). In panel (a), the experimental data are represented by filled symbols (listed in the figure key). For each $\alpha$, a power law $H \propto U^\gamma$ was fitted to best match the experimental data points, the horizontal lines indicating $H_{if}$, and the inset showing the exponent $\gamma$. Panel (b) displays the data in log scales, with the fitting curve being accurately described by a power law ($c$ $=$ 1.07, $\gamma$ $=$ 2.1, and coefficient of determination $R^2 = 0.9982$).}
	\label{fig:flow_vs_bedheight_all_inset.pdf}
\end{figure} 

In order to inquire further into the behavior of the bed height $H$ for different bonding angles $\alpha$, we plotted $H$ as a function of the superficial velocity $U$ and parameterized by $\alpha$, which is shown in Fig. \ref{fig:flow_vs_bedheight_all_inset.pdf}a in linear scales and dimensional form, and in Fig. \ref{fig:flow_vs_bedheight_all_inset.pdf}b in log-log scales and dimensionless form. From Fig. \ref{fig:flow_vs_bedheight_all_inset.pdf}b, we observe that, irrespective of the type of trio, the height follows a power-law relation,

\begin{equation}
	\frac{H}{H_{if}} = c \left( \frac{U}{U_{if}}\right) ^{\gamma} \,\,,
	\label{eq:H}
\end{equation}

\noindent implying that system would probably behave similarly across a wide range of scales (i.e., very-narrow SLFBs of different absolute sizes), but the relatively small range of variation in $U$ hiders us from asserting that. In Eq. \ref{eq:H}, $c$ and $\gamma$ are constants, for which we found 1.07 and 2.1, respectively.

Finally, we inquire further into the plug structures appearing in the fluidized (Fd) regime. Table \ref{tab_plug_length} presents the mean values of the plug length $\lambda / D$ and the corresponding standard deviations $\sigma_{\lambda} / D$  for different angles $\alpha$ and water velocities $U/U_{if}$. From this table, we observe that  $\lambda / D$ is of the order of unity, as found previously for loose \cite{Oliveira} and bonded spheres \cite{Cunez4}, with maximum values (still around unity) reached when $\alpha$ $=$ 150$^{\circ}$ and minimum values when $\alpha$ $=$ 60$^{\circ}$ and 90$^{\circ}$. In principle, lower values of $\lambda / D$ imply lower local concentrations of grains and, therefore, better mixture between water and solids (and, thus, higher rates of heat and mass transfers). However, differences between minimum and maximum values of the mean plug length $\lambda / D$ are rather small, and the previous conclusions must be taken with caution (a graphic of Tab. \ref{tab_plug_length} is available in the supplementary material).

\begin{table}[h]
	\centering
	\caption{For the fluidized regime (Fd), mean values of the plug length $\lambda / D$ and the corresponding standard deviations $\sigma_{\lambda} / D$  for different angles $\alpha$ and water velocities $U/U_{if}$.}
	\label{tab_plug_length}
	\color{black} 
	\begin{tabular}{c c c c c c c}
		\hline
		\hline
		$\alpha$ ($^{\circ}$) & $U/U_{if}$ & $\lambda / D$ & $\sigma_{\lambda}/D$\\
		\hline
		60 & 1.76 & 0.70 & 0.20\\
		60 & 2.07 & 0.64 & 0.23\\
		60 & 2.36 & 0.60 & 0.18\\
		90 & 1.49 & 0.74 & 0.23\\
		90 & 1.75 & 0.67 & 0.20\\
		90 & 1.99 & 0.62 & 0.61\\
		120 & 1.66 & 0.74 & 0.28\\
		150 & 1.66 & 0.78 & 0.61\\
		150 & 1.94 & 0.67 & 0.21\\
		180 & 1.49 & 0.80 & 0.29\\
		180 & 1.75 & 0.71 & 0.23\\
		180 & 1.99 & 0.65 & 0.45\\
		\hline
		\hline
	\end{tabular}
\end{table}

For the regimes in which the fluidization was hindered in some sort, namely the CP, Cg and GT regimes, we measured the durations of the observed regime and the respective initial and final times in the case of the GT regime. We show next the ensemble-averages for each experimental condition, and present in Appendix \ref{appendix} (Tabs. \ref{table:tests_CP}, \ref{table:tests_Cg} to \ref{table:tests_GT}) the data for each experimental test run in which those regimes appeared.

Tables \ref{table:averages_CP} and \ref{table:averages_Cg} show the angle between particles, $\alpha$, dimensionless superficial velocity, $U/U_{if}$, ensemble-averaged duration, $\left< \Delta t \right>$, and maximum duration, $\Delta t_{max}$, for the clogged-plug (CP) and clogging (Cg) regimes, respectively. The averages were computed for each tested condition, that is, over all events taking place in all test runs of a given experimental condition ($U$ and $\alpha$ fixed). For the CP regime, we observe from Tab. \ref{table:averages_CP} that in most of cases the highest values of $\left< \Delta t \right>$ and $\Delta t_{max}$ occur for the smallest fluid velocities, and values are greater when $\alpha$ $=$ 60$^{\circ}$. This means that the CP regime appears and lasts more time when $\alpha$ $=$ 60$^{\circ}$ and $U/U_{if}$ $\approx$ 1,, i.e., the highest probability of finding CPs occurs under these conditions. This corroborates the map presented in Fig. \ref{fig:regime_map}, and the corresponding discussion. On the other hand, for the Cg regime Tab. \ref{table:averages_Cg} shows that it is for $\alpha$ $=$ 120$^{\circ}$ and 180$^{\circ}$ that the highest durations are found, when $U/U_{if}$ $\approx$ 1. We would like to note that the Cg regime does not appear in the map of Fig. \ref{fig:regime_map} because another regime was dominant (it coexisted with other regime that prevailed). We also note that sometimes $U/U_{if}$ $<$ 1 because of the uncertainties intrinsic in measuring $U_{if}$ in a system of confined bounded particles (identified by image processing, as described in Section \ref{sec:exp_setup}). Finally, Tab. \ref{table:averages_GT} shows that it is under low velocities and for $\alpha$ $=$ 150$^{\circ}$ and 180$^{\circ}$ that the GT regime appears, and that it forms faster (lower values of $\left< \tau \right>$) when $\alpha$ $=$ 180$^{\circ}$ (also corroborating the map presented in Fig. \ref{fig:regime_map} and the corresponding discussion, the data for $\alpha$ $=$ 90$^{\circ}$ not appearing in Fig. \ref{fig:regime_map} because it coexisted with other regime that prevailed).


\begin{longtable}{P{1cm}P{1.5cm}P{1.5cm}P{2.5cm}} 
	\caption{Angle between particles, $\alpha$, dimensionless superficial velocity, $U/U_{if}$, ensemble-averaged duration, $\left< \Delta t \right>$, and maximum duration, $\Delta t_{max}$, for the clogged-plug (CP) regime. Averages were computed over all events taking place in all test runs of a given case.} \label{table:averages_CP} \\
	\toprule
	$\alpha$ ($^{\circ}$) & $U/U_{if}$ & $\left< \Delta t \right>$ (s) & $\Delta t_{max}$ (s)\\
	\midrule
	\endfirsthead
	\caption[]{(continued) Angle between particles, $\alpha$, dimensionless superficial velocity, $U/U_{if}$, ensemble-averaged duration, $\left< \Delta t \right>$, and maximum duration, $\Delta t_{max}$, for the clogged-plug (CP) regime. Averages were computed over all events taking place in all test runs of a given case.}\\
	\toprule
	$\alpha$ ($^{\circ}$) & $U/U_{if}$ & $\left< \Delta t \right>$ (s) & $\Delta t_{max}$ (s)\\
	\midrule
	\endhead
	\midrule \multicolumn{4}{r}{{}} \\
	\endfoot
	\bottomrule
	\endlastfoot
	
	60 & 1.00 & 42 & 55\\ 
	60 & 1.06 & 17 & 26\\ 
	60 & 1.12 & 13 & 16\\ 
	60 & 1.18 & 10 & 17\\ 
	60 & 1.24 & 7 & 13\\ 
	60 & 1.30 & 7 & 15\\ 
	60 & 1.35 & 7 & 9\\ 
	60 & 1.42 & 5 & 9\\ 
	60 & 1.47 & 6 & 10\\ 
	90 & 0.90 & 17 & 17\\ 
	90 & 0.95 & 15 & 22\\ 
	90 & 1.00 & 10 & 16\\ 
	90 & 1.05 & 10 & 17\\ 
	90 & 1.10 & 9 & 11\\ 
	90 & 1.15 & 14 & 41\\ 
	90 & 1.20 & 10 & 33\\ 
	90 & 1.25 & 6 & 8\\ 
	90 & 1.49 & 4 & 4\\ 
	120 & 1.00 & 26 & 53\\ 
	120 & 1.11 & 18 & 23\\ 
	120 & 1.16 & 6 & 8\\ 
	120 & 1.22 & 6 & 8\\ 
	120 & 1.27 & 8 & 14\\ 
	120 & 1.33 & 9 & 18\\ 
	120 & 1.38 & 6 & 9\\ 
	150 & 1.00 & 4 & 4\\ 
	150 & 1.05 & 16 & 24\\ 
	150 & 1.11 & 5 & 6\\ 
	150 & 1.16 & 5 & 8\\ 
	150 & 1.22 & 3 & 3\\ 
	150 & 1.27 & 4 & 4\\ 
	150 & 1.33 & 13 & 29\\ 
	150 & 1.38 & 5 & 7\\ 
	180 & 1.00 & 4 & 4\\ 
	180 & 1.05 & 14 & 21\\ 
	180 & 1.10 & 6 & 8\\ 
	180 & 1.15 & 7 & 9\\ 
	180 & 1.20 & 8 & 12\\ 
	180 & 1.25 & 4 & 7\\ 
	180 & 1.49 & 8 & 10\\
\end{longtable}


\begin{longtable}{P{1cm}P{1.5cm}P{1.5cm}P{2.5cm}}
	\caption{Angle between particles, $\alpha$, dimensionless superficial velocity, $U/U_{if}$, ensemble-averaged duration, $\left< \Delta t \right>$, and maximum duration,  $\Delta t_{max}$, for the clogging (Cg) regime. Averages were computed over all events taking place in all test runs of a given case.}\label{table:averages_Cg}\\
	\toprule
	$\alpha$ ($^{\circ}$) & $U/U_{if}$ & $\left< \Delta t \right>$ (s) & $\Delta t_{max}$ (s)\\
	\midrule
	\endfirsthead
	\caption[]{(continued) Angle between particles, $\alpha$, dimensionless superficial velocity, $U/U_{if}$, ensemble-averaged duration, $\left< \Delta t \right>$, and maximum duration,  $\Delta t_{max}$, for the clogging (Cg) regime. Averages were computed over all events taking place in all test runs of a given case.}\\
	\toprule
	$\alpha$ ($^{\circ}$) & $U/U_{if}$ & $\left< \Delta t \right>$ (s) & $\Delta t_{max}$ (s)\\
	\midrule 
	\endhead
	\midrule \multicolumn{4}{r}{{}} \\
	\endfoot
	\bottomrule
	\endlastfoot
	60 & 1.06 & 126 & 202\\
	120 & 0.94 & 300 & 300\\ 
	120 & 1.00 & 300 & 300\\ 
	120 & 1.05 & 157 & 300\\ 
	120 & 1.11 & 111 & 300\\ 
	120 & 1.16 & 190 & 300\\ 
	120 & 1.22 & 211 & 300\\ 
	120 & 1.27 & 264 & 264\\ 
	120 & 1.33 & 17 & 24\\ 
	150 & 0.94 & 155 & 155\\ 
	150 & 1.00 & 270 & 270\\ 
	150 & 1.05 & 213 & 300\\ 
	150 & 1.11 & 98 & 146\\ 
	150 & 1.16 & 44 & 48\\ 
	150 & 1.22 & 262 & 287\\ 
	150 & 1.27 & 208 & 283\\ 
	150 & 1.33 & 9 & 9\\ 
	150 & 1.38 & 30 & 30\\ 
	180 & 0.80 & 300 & 300\\ 
	180 & 1.00 & 267 & 267\\ 
	180 & 1.10 & 247 & 247\\ 
	180 & 1.15 & 16 & 16\\ 
	180 & 1.25 & 18 & 18\\ 
\end{longtable}


\begin{longtable}{P{1cm}P{1.5cm}P{1.5cm}P{2.5cm}}
	\caption{Angle between particles, $\alpha$, dimensionless superficial velocity, $U/U_{if}$, ensemble-averaged time to reach the regime, $\left< \tau \right>$, and maximum time for reaching the regime, $\tau_{max}$, for the glass-transition (GT) regime. Averages were computed over all events taking place in all test runs of a given case.}\label{table:averages_GT}\\
	\toprule
	$\alpha$ ($^{\circ}$) & $U/U_{if}$ & $\left< \tau \right>$ (s) & $\tau_{max}$ (s)\\
	\midrule
	\endfirsthead
	\caption[]{(continued) Angle between particles, $\alpha$, dimensionless superficial velocity, $U/U_{if}$, ensemble-averaged time to reach the regime, $\left< \tau \right>$, and maximum time for reaching the regime, $\tau_{max}$, for the glass-transition (GT) regime. Averages were computed over all events taking place in all test runs of a given case.}\\
	\toprule
	$\alpha$ ($^{\circ}$) & $U/U_{if}$ & $\left< \tau \right>$ (s) & $\tau_{max}$ (s)\\ 
	\midrule
	\endhead
	\midrule \multicolumn{4}{r}{{}} \\
	\endfoot
	\bottomrule
	\endlastfoot
	
	90 & 0.95 & 33 & 33\\ 
	150 & 0.94 & 25 & 25\\ 
	150 & 1.00 & 116 & 117\\ 
	150 & 1.05 & 92 & 155\\ 
	180 & 1.00 & 10 & 13\\ 
	180 & 1.05 & 51 & 118\\ 
	180 & 1.10 & 21 & 48\\ 
	180 & 1.15 & 80 & 142\\ 
	180 & 1.20 & 54 & 86\\ 
	180 & 1.25 & 11 & 11\\
\end{longtable}

\begin{figure}[ht]
	\centering
	\includegraphics[width=0.8\columnwidth]{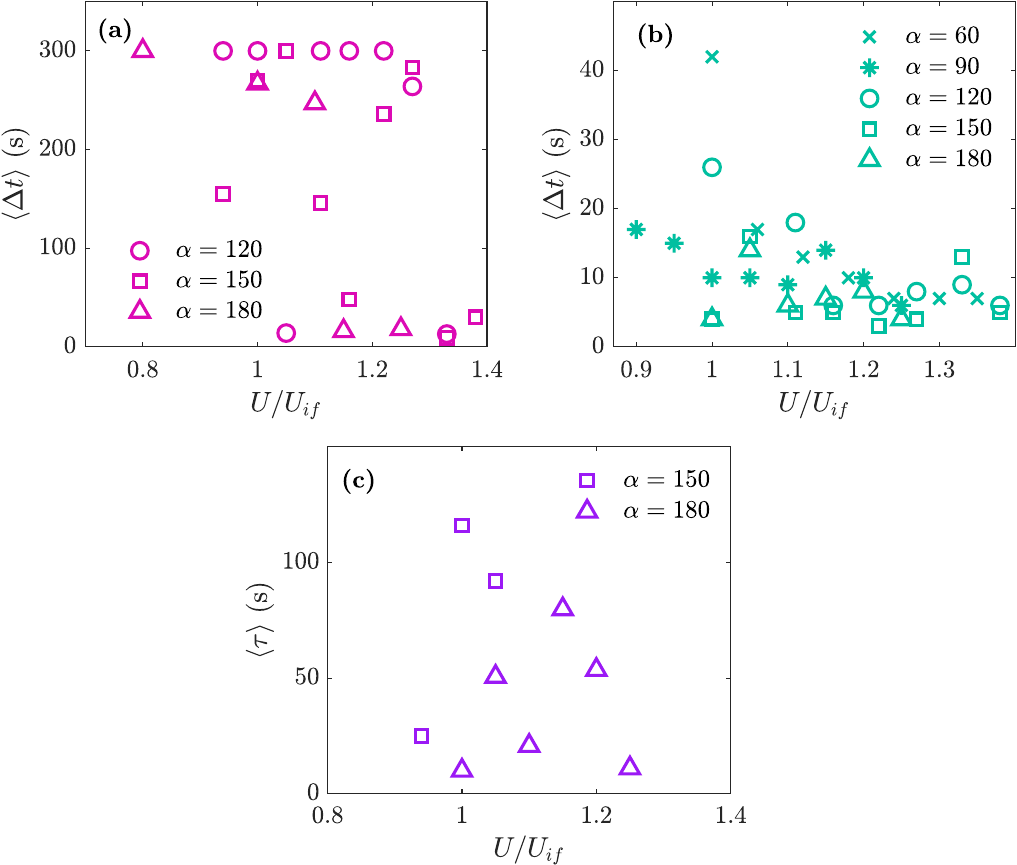}
	\caption{Mean durations $\left< \Delta t \right>$ and times $\left< \tau \right>$ as functions of the fluid velocity $U/U_{if}$, parameterized by the bonding angle $\alpha$. The figures correspond to data listed in Tabs. \ref{table:averages_CP} to \ref{table:averages_GT}: (a) Clogged-plug (CP) regime; (b) Clogging (Cg) regime; (c) Glass-transition (GT) regime.}
	\label{fig:mean_times}
\end{figure}

Finally, Fig. \ref{fig:mean_times} presents the mean durations $\left< \Delta t \right>$ and times $\left< \tau \right>$ listed in Tabs. \ref{table:averages_CP} to \ref{table:averages_GT}, as functions of the fluid velocity $U/U_{if}$ and parameterized by the bonding angle $\alpha$. From Fig. \ref{fig:mean_times}a, we observe that the CP regime appears only for $120^{\circ}$ $\leq$ $\alpha$ $\leq$ $180^{\circ}$, but with no clear tendency of $\left< \Delta t \right>$ with $U$. Fig. \ref{fig:mean_times}b, on the other hand, shows that the Cg regime appears for all values of $\alpha$, but becomes less intense (lower values of $\left< \Delta t \right>$) as $U$ increases. For the GT regime (Fig. \ref{fig:mean_times}c), we observe clearly the lower times for reaching the regime when $\alpha$ $=$ 180$^{\circ}$.

\subsection{Microscopic scale}

\begin{figure}[ht]
	\centering
	\includegraphics[width=0.75\columnwidth]{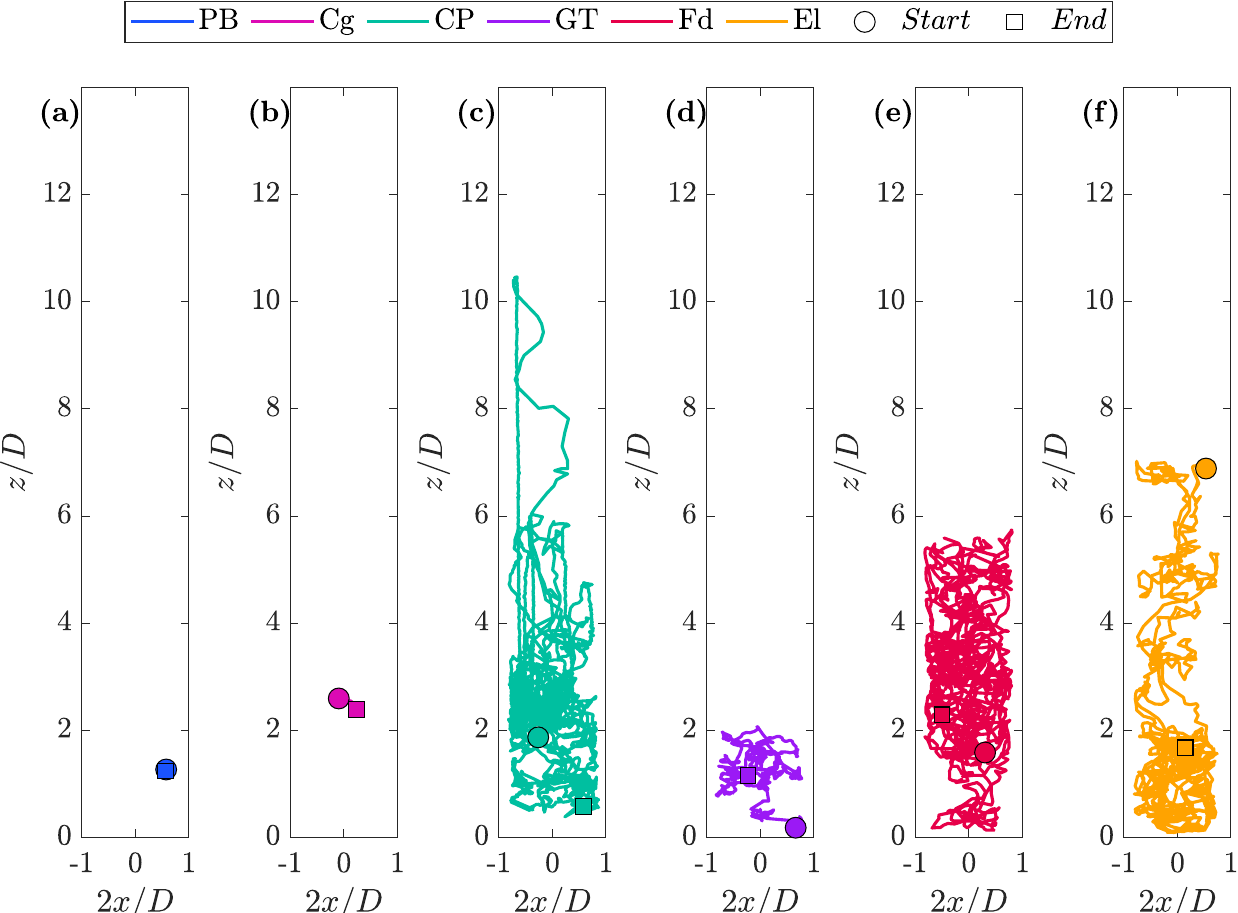}
	\caption{Examples of trajectories for the different regimes observed: (a) packed bed (PB), (b) clogging (Cg), (c) clogged-plug (CP), (d) glass transition (GT), (e) fluidized bed (Fb), and (f) elutriation (El). The time intervals for which the particles were tracked are 300 s for panels a-d, 66 s for panel e, and 48 s for panel f. $z$ is the vertical coordinate and $x$ is the horizontal coordinate (with origin in the center of the tube).}
	\label{fig:trajectories}
\end{figure}

We now investigate the motion of individual particles for the different regimes observed in Subsection \ref{subsec:macroscopic}. In terms of typical trajectories, Fig. \ref{fig:trajectories} shows some examples for one single agglomerate, from which we can observe an absence of motion for the PB regime (Fig. \ref{fig:trajectories}a), and virtually no motion for he Cg regime (Fig. \ref{fig:trajectories}b, after the clogging process has taken place). For the GT regime (Fig. \ref{fig:trajectories}d, shown from the beginning to the end of the glass transition process), the motion of the particle is relatively short, being confined in lower regions of the tube while the amorphous static structure is being formed. For the CP regime (Fig. \ref{fig:trajectories}c), the motion spans over the tube cross section and a large portion of its height. From some time instant on, the particle moves vertically without any horizontal motion: it remains trapped in the clog as the latter moves upward while decaying (by the free fall of particles in its lower part). Finally, in the Fd regime (Fig. \ref{fig:trajectories}e), the particle motion spans over the tube cross section and a large portion of its height, at all times, while in the El regime (Fig. \ref{fig:trajectories}f) the particle remains for a while in the lower part of the tube, after which it is entrained further downstream.

\begin{figure}[ht]
	\centering
	\includegraphics[width=0.99\columnwidth]{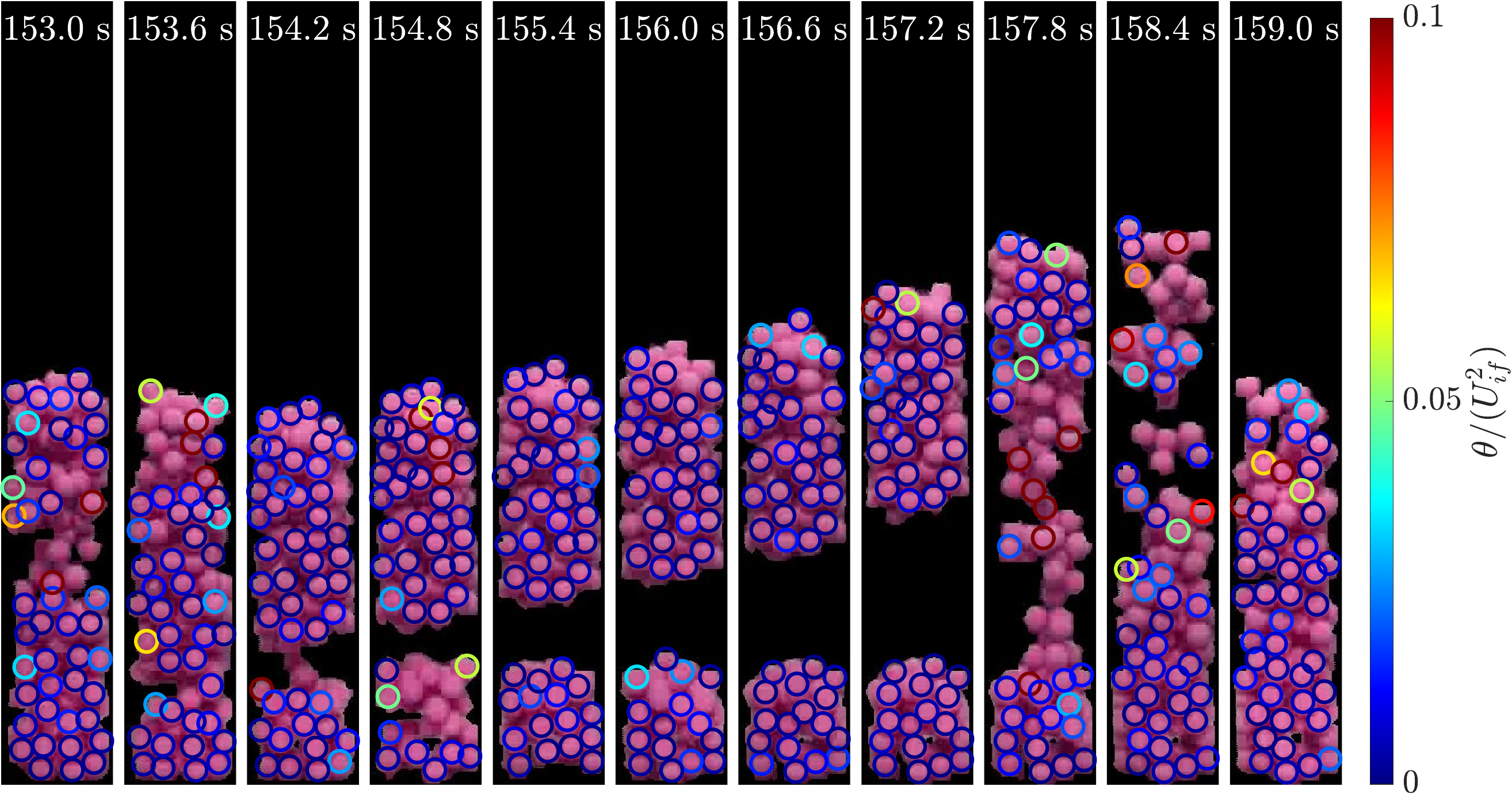}
	\caption{Instantaneous granular temperatures of some particles superposed with snapshots showing the bed, for $U$ $=$ 0.104 m/s and $\alpha$ $=$ 90$^{\circ}$. The time instants are shown on the top of each panel, and the color of each identified particle corresponds to the granular temperature shown on the colorbar on the right of the figure. Values are normalized by $U_{if}^2$.}
	\label{fig:snap_temp_granular}
\end{figure}

In order to analyze the motion of grains as an ensemble, we computed their individual granular temperature, which is a measure of the fluctuation of particles. For that, we computed ensemble averages as in Eq. \ref{temp_def} \cite{Andreotti_2, Campbell, Peng2},

\begin{equation}
	\theta(t) = \frac{1}{2N}\sum_{i=1}^{N}\left({u}^2_{i}+{v}^2_{i}\right) \,\,,
	\label{temp_def}
\end{equation}

\begin{figure}[ht]
	\centering
	\includegraphics[width=0.65\columnwidth]{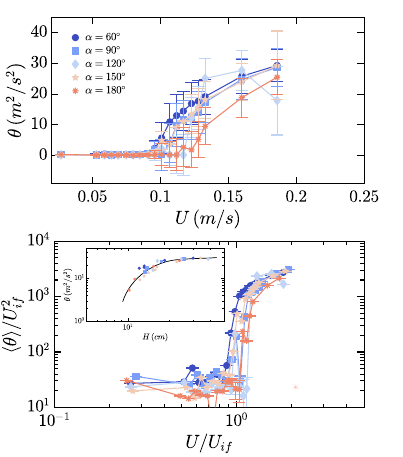}
	\caption{Space-time averages of the granular temperature, $\left< \theta \right>$, as a function of the dimensional ($U$, panel a) and dimensionless ($U/U_{if}$, panel b) velocities, parameterized by the bonding angle $\alpha$. The symbols are listed in the figure key, the inset in panel (b) shows how the granular temperature varies with $H$, and in panel (b) $\left< \theta \right>$ is normalized by $U_{if}^2$.}
	\label{fig:temp_granular}
\end{figure} 

\noindent where $u_i$ and $v_i$ are, respectively, the horizontal and vertical velocity fluctuations of the $i^{\rm th}$ particle, and $N$ is the number of particles used in the computations (those for which we have optical access, i.e., those in contact with the tube wall). An example of instantaneous values of granular temperature can be seen in Fig. \ref{fig:snap_temp_granular}, in which snapshots of the bed are placed side by side, superposed with the values of $\theta$ for some of the spheres.

\begin{table}[h]
	\centering
	\caption{Space-time averages of granular temperatures for each angle $\alpha$ (subscripts) and superficial velocity $U$. The values for loose spheres $\left< \theta_{loose} \right>$ are shown for reference.}
	\label{tab:tab2}
	\begin{tabular}{m{1.5cm} m{1.5cm} m{1.5cm} m{1.5cm} m{1.5cm} m{1.5cm} m{1.5cm}}
		\hline
		\hline
		U & $\left< \theta_{loose} \right>$ & $\left< \theta_{\alpha,60} \right>$ & $\left< \theta_{\alpha,90} \right>$ & $\left< \theta_{\alpha,120} \right>$ & $\left< \theta_{\alpha,150} \right>$ & $\left< \theta_{\alpha,180} \right>$\\
		(cm/s) & (cm/s)$^2$ & (cm/s)$^2$ & (cm/s)$^2$ & (cm/s)$^2$ & (cm/s)$^2$ & (cm/s)$^2$ \\
		\hline
		2.7 & 0.78 & 0.29 & 0.34 & 0.25 & 0.20 & 0.36 \\
		5.5 & 0.83 & 0.31 & 0.24 & 0.30 & 0.23 & 0.19 \\
		6.0 & 1.06 & 0.55 & 0.36 & 0.20 & 0.15 & 0.17 \\
		6.6 & 1.09 & 0.30 & 0.23 & 0.20 & 0.35 & 0.23 \\
		7.1 & 0.90 & 0.30 & 0.32 & 0.20 & 0.23 & 0.20 \\
		7.7 & 0.73 & 0.38 & 0.42 & 0.28 & 0.30 & 0.19 \\
		8.2 & 0.88 & 0.46 & 0.36 & 0.22 & 0.27 & 0.05 \\
		8.8 & 1.02 & 0.28 & 0.28 & 0.25 & 0.42 & 0.23 \\
		9.3 & - & 0.61 & 0.69 & 0.28 & 0.39 & 0.23 \\
		9.9 & - & 2.36 & 1.78 & 0.22 & 1.04 & 0.34 \\
		10.4 & - & 5.74 & 0.27 & 2.27 & 4.93 & 0.33 \\
		11.0 & 6.53 & 10.96 & 4.02 & 0.25 & 3.61 & 0.25 \\
		11.5 & - & 13.01 & 9.83 & 0.17 & 9.22 & 0.25 \\
		12.1 & - & 14.49 & 11.61 & 0.23 & 11.18 & 2.75 \\
		12.6 & - & 16.93 & 12.20 & 12.33 & 13.50 & 1.88 \\
		13.2 & - & 17.65 & 14.11 & 15.52 & 15.62 & 5.27 \\
		13.7 & 15.25 & 19.18 & 17.37 & 25.10 & 18.16 & 9.54 \\
		16.4 & 20.39 & 25.81 & 24.78 & 27.71 & 24.00 & 18.84 \\
		19.2 & 24.29 & 29.32 & 28.70 & 17.92 & 29.32 & 25.49 \\
		\hline
		\hline
	\end{tabular}
\end{table}

The time average of the ensemble-averaged granular temperature (corresponding to space-time averages), $\left< \theta \right>$, can be then computed by Eq. \ref{temp_time_avg} \cite{Goldhirsch, Maranic},

\begin{equation}
	\left< \theta \right> = \frac{1}{T}\sum_{t=0}^{T} \theta \,\,,
	\label{temp_time_avg}
\end{equation}

\noindent where $T$ is the total duration of the experiment. Figure \ref{fig:temp_granular} shows how $\left< \theta \right>$ varies with $U$ (Fig. \ref{fig:temp_granular}a) and $U/U_{if}$ (Fig. \ref{fig:temp_granular}b) for all the bonding angles used. Comparing with Fig. \ref{fig:regime_map}, we basically observe very low values of $\left< \theta \right>$ in the PB regime, and high values for the Fd and El regimes, with intermediate values for the CP, Cg and GT regimes. The inset in Fig. \ref{fig:temp_granular}b shows the variation of the granular temperature with the bed height, parameterized by the bonding angle, and we cannot observe a clear dependence on the latter (although it can exist one, symbols for 60$^{\circ}$ $\leq$ $\alpha$ $\leq$ 120$^{\circ}$ having slightly  higher values of $\theta$).

From Fig. \ref{fig:temp_granular}, it is difficult to distinguish how $\left< \theta \right>$ varies with $\alpha$, and for that reason we list the values for each $\alpha$ in Tab. \ref{tab:tab2} (the corresponding standard deviations are listed in Tab. \ref{tab:tab2_std} in Appendix \ref{appendix}). In the Tab. \ref{tab:tab2}, we included the value obtained for loose spheres (not bonded) for reference. We clearly observe that $\left< \theta \right>$ is higher for $\alpha$ $=$ 60$^{\circ}$ and $\alpha$ $=$ 90$^{\circ}$, indicating better fluidization for these bonding angles. This corroborates our conclusions on optimal fluidization conditions presented in Subsection \ref{subsec:macroscopic}.

In summary, our results show that, depending on the bonding angle and fluid velocity, five different regimes appear for beds consisting of trios of spheres, and we propose a classification map that organizes the results. The proposed map can be used as a guide for selecting the fluid velocities for a given configuration in order to keep the bed fluidized at all times. Therefore, it helps in the design of fluidized beds and their operation. In addition, the results indicate (based on the fluidization range, bed height, and particle fluctuations) that the optimal angle (among those tested) is 90$^{\circ}$.

\section{CONCLUSIONS}
\label{sec:conclusions}

In this paper, we inquired into the deterioration of very-narrow fluidized beds of bonded particles by the formation of large aggregates, sometimes static. For that, we carried out experiments in which the particles consisted of trios of spheres fluidized in water, and we varied the bonding angle $\alpha$ and the flow velocity $U$. We showed that \corr{regimes not observed in previous works with bonded grains \cite{Cunez4}} can appear, namely packed beds (PB), persistent fluidization with the presence of plugs (Fd), clogging (Cg), elutriation (El), a mix of cloggind and plug (a moving clog) that we named clogged plug (CP), and an static amorphous structure that we named glass transition (GT). We proposed a classification map in the $\alpha$ -- $U/U_{if}$ space, in which the patterns were well organized. It basically shows that the PB regime occurs for velocities lower than that for incipient fluidization $U_{if}$, and that for larger velocities we have, in sequence of increasing velocity, the Fd and El regimes, for all values of $\alpha$. For intermediate values of $U$ we find the CP regime for all values of $\alpha$, and the Cg and GT regimes for $\alpha$ $\geq$ 120$^{\circ}$.  Of all regimes, the PB,  Cg and GT are static, meaning that the fluidized bed no longer fills its role. We measured the characteristic durations of the CP and Cg regimes, which showed that the former is more intense for lower values and the latter to higher values of $\alpha$, and the characteristic time for reaching the GT regime, which showed that it forms faster when $\alpha$ $=$ 180$^{\circ}$. We also measured the fluctuation of individual particles in terms of granular temperature $\theta$ and its space-time average $\left< \theta \right>$, from which we show that higher $\left< \theta \right>$ occurs for $\alpha$ $=$ 60$^{\circ}$ and $\alpha$ $=$ 90$^{\circ}$. Based on data from both the macroscopic (bed) and microscopic (grain) scales, we find that agglomerates of spheres bonded with $\alpha$ $=$ 90$^{\circ}$ remain suspended and fluctuating over larger ranges of $U$ and, thus, have the best fluidization properties among trios. Different from previous works \cite{Cunez4}, we showed that the bonding angle influences significantly the behavior of the bed, with different regimes appearing depending on $\alpha$ and $U$. Despite the simplifications made in the experiments (agglomerates of three spheres forming planar angles), these results bring new insights into problems involving cohesive particles fluidized by water, such as happens in bioreactors for wastewater treatment.

\section*{AUTHOR DECLARATIONS}
\noindent \textbf{Conflict of Interest}

The authors have no conflicts to disclose

\section*{SUPPLEMENTARY MATERIAL}
See the supplementary material for additional figures of our results.

\section*{DATA AVAILABILITY}
The data that support the findings of this study are openly available in Mendeley Data at https://doi.org/10.17632/s3r8g8f3t7.1.

\begin{acknowledgments}
The authors are grateful to FAPESP (Grant Nos. 2018/14981-7, 2020/00221-0 and 2022/01758-3) and to CNPq (Grant No. 405512/2022-8) for the financial support provided. We acknowledge support by the Open Access Publication Fund of the University of Duisburg-Essen.
\end{acknowledgments}

\clearpage

\appendix
\section{Data by test run}
\label{appendix}

This Appendix brings tables with the initial, final, and duration times for each test run.
 

\begin{longtable}{P{1cm}P{2cm}P{1cm}P{2cm}P{2.5cm}}
	\caption{Angle between particles, $\alpha$, dimensionless superficial velocity, $U/U_{if}$, label (number) of the test run (set), ensemble-averaged duration for each set, $\left< \Delta t \right>_{set}$, and maximum duration for each set, $\Delta t_{max, set}$, for the clogged-plug regime.}\label{table:tests_CP} \\
	\toprule
	$\alpha$ ($^{\circ}$) & $U/U_{if}$ & Set & $\left< \Delta t \right>_{set}$ (s)& $\Delta t_{max, set}$ (s)\\
	\midrule
	\endfirsthead
	\caption[]{(continued) Angle between particles, $\alpha$, dimensionless superficial velocity, $U/U_{if}$, label (number) of the test run (set), ensemble-averaged duration for each set, $\left< \Delta t \right>_{set}$, and maximum duration for each set, $\Delta t_{max, set}$, for the clogged-plug regime.} \\
	\toprule
	$\alpha$ ($^{\circ}$) & $U/U_{if}$ & Set & $\left< \Delta t \right>_{set}$ (s) & $\Delta t_{max, set}$ (s)\\
	\midrule
	\endhead
	\midrule \multicolumn{5}{r}{{}} \\
	\endfoot
	\bottomrule
	\endlastfoot
	
	60 & 1.00 & 1 & 55.25 & 137 \\
	60 & 1.00 & 3 & 29.67 & 64 \\
	60 & 1.06 & 1 & 25.50 & 96 \\
	60 & 1.06 & 2 & 20.67 & 44 \\
	60 & 1.06 & 3 & 6.50 & 8 \\
	60 & 1.06 & 4 & 23.00 & 61 \\
	60 & 1.06 & 5 & 9.50 & 12 \\
	60 & 1.12 & 1 & 11.38 & 23 \\
	60 & 1.12 & 2 & 9.00 & 22 \\
	60 & 1.12 & 3 & 13.40 & 22 \\
	60 & 1.12 & 4 & 15.86 & 50 \\
	60 & 1.12 & 5 & 14.50 & 35 \\
	60 & 1.18 & 1 & 7.33 & 10 \\
	60 & 1.18 & 2 & 4.38 & 9 \\
	60 & 1.18 & 3 & 17.00 & 26 \\
	60 & 1.18 & 4 & 12.75 & 20 \\
	60 & 1.18 & 5 & 8.88 & 26 \\
	60 & 1.24 & 1 & 4.86 & 9 \\
	60 & 1.24 & 2 & 4.10 & 11 \\
	60 & 1.24 & 3 & 4.00 & 4 \\
	60 & 1.24 & 4 & 9.50 & 12 \\
	60 & 1.24 & 5 & 13.25 & 23 \\
	60 & 1.30 & 1 & 4.00 & 7 \\
	60 & 1.30 & 2 & 6.00 & 14 \\
	60 & 1.30 & 3 & 4.33 & 6 \\
	60 & 1.30 & 4 & 6.00 & 11 \\
	60 & 1.30 & 5 & 14.75 & 34 \\
	60 & 1.35 & 1 & 3.50 & 4 \\
	60 & 1.35 & 2 & 6.33 & 9 \\
	60 & 1.35 & 4 & 9.33 & 13 \\
	60 & 1.35 & 5 & 8.00 & 13 \\
	60 & 1.42 & 1 & 3.60 & 7 \\
	60 & 1.42 & 2 & 8.50 & 15 \\
	60 & 1.42 & 3 & 4.50 & 6 \\
	60 & 1.42 & 4 & 2.50 & 3 \\
	60 & 1.42 & 5 & 4.67 & 8 \\
	60 & 1.47 & 2 & 3.50 & 4 \\
	60 & 1.47 & 3 & 10.00 & 10 \\
	60 & 1.47 & 5 & 3.67 & 6 \\
	90 & 0.90 & 5 & 16.50 & 24 \\
	90 & 0.95 & 3 & 7.71 & 27 \\
	90 & 0.95 & 4 & 22.00 & 22 \\
	90 & 1.00 & 1 & 15.75 & 81 \\
	90 & 1.00 & 3 & 6.75 & 13 \\
	90 & 1.00 & 4 & 4.88 & 8 \\
	90 & 1.00 & 5 & 12.80 & 41 \\
	90 & 1.05 & 1 & 4.83 & 11 \\
	90 & 1.05 & 2 & 10.20 & 41 \\
	90 & 1.05 & 3 & 16.60 & 57 \\
	90 & 1.05 & 4 & 9.00 & 14 \\
	90 & 1.05 & 5 & 11.00 & 28 \\
	90 & 1.10 & 1 & 7.83 & 16 \\
	90 & 1.10 & 2 & 8.00 & 12 \\
	90 & 1.10 & 3 & 11.00 & 22 \\
	90 & 1.10 & 4 & 6.50 & 9 \\
	90 & 1.10 & 5 & 11.20 & 13 \\
	90 & 1.15 & 1 & 5.67 & 10 \\
	90 & 1.15 & 2 & 7.00 & 10 \\
	90 & 1.15 & 3 & 12.50 & 52 \\
	90 & 1.15 & 4 & 40.67 & 111 \\
	90 & 1.15 & 5 & 4.00 & 6 \\
	90 & 1.20 & 1 & 5.57 & 21 \\
	90 & 1.20 & 2 & 4.20 & 5 \\
	90 & 1.20 & 3 & 4.00 & 8 \\
	90 & 1.20 & 4 & 5.10 & 13 \\
	90 & 1.20 & 5 & 33.00 & 64 \\
	90 & 1.25 & 1 & 5.33 & 10 \\
	90 & 1.25 & 2 & 3.29 & 6 \\
	90 & 1.25 & 3 & 7.80 & 17 \\
	90 & 1.25 & 4 & 8.00 & 16 \\
	90 & 1.25 & 5 & 7.00 & 13 \\
	90 & 1.49 & 2 & 4.00 & 4 \\
	120 & 1.00 & 4 & 53.00 & 53 \\
	120 & 1.11 & 2 & 20.00 & 52 \\
	120 & 1.11 & 3 & 9.67 & 14 \\
	120 & 1.11 & 4 & 23.00 & 44 \\
	120 & 1.16 & 3 & 8.33 & 18 \\
	120 & 1.16 & 4 & 4.00 & 4 \\
	120 & 1.22 & 2 & 3.50 & 7 \\
	120 & 1.22 & 4 & 7.50 & 10 \\
	120 & 1.27 & 1 & 8.50 & 13 \\
	120 & 1.27 & 2 & 8.50 & 18 \\
	120 & 1.27 & 3 & 2.60 & 3 \\
	120 & 1.27 & 4 & 14.00 & 24 \\
	120 & 1.33 & 1 & 3.50 & 6 \\
	120 & 1.33 & 3 & 18.00 & 18 \\
	120 & 1.33 & 4 & 6.00 & 12 \\
	120 & 1.38 & 3 & 5.50 & 13 \\
	120 & 1.38 & 4 & 8.50 & 14 \\
	120 & 1.38 & 5 & 3.00 & 3 \\
	150 & 1.00 & 3 & 4.00 & 5 \\
	150 & 1.05 & 1 & 12.50 & 52 \\
	150 & 1.05 & 3 & 11.00 & 11 \\
	150 & 1.05 & 5 & 23.50 & 39 \\
	150 & 1.11 & 1 & 5.75 & 11 \\
	150 & 1.11 & 3 & 5.00 & 7 \\
	150 & 1.16 & 1 & 3.80 & 6 \\
	150 & 1.16 & 3 & 4.67 & 6 \\
	150 & 1.16 & 5 & 8.00 & 8 \\
	150 & 1.22 & 1 & 3.29 & 6 \\
	150 & 1.22 & 3 & 3.00 & 3 \\
	150 & 1.27 & 3 & 4.00 & 6 \\
	150 & 1.33 & 3 & 4.25 & 7 \\
	150 & 1.33 & 4 & 5.50 & 6 \\
	150 & 1.33 & 5 & 29.00 & 29 \\
	150 & 1.38 & 3 & 4.00 & 4 \\
	150 & 1.38 & 4 & 7.00 & 7 \\
	150 & 1.38 & 5 & 4.00 & 4 \\
	180 & 1.00 & 3 & 4.00 & 4 \\
	180 & 1.05 & 3 & 7.00 & 7 \\
	180 & 1.05 & 5 & 21.00 & 21 \\
	180 & 1.10 & 2 & 4.50 & 5 \\
	180 & 1.10 & 5 & 8.25 & 12 \\
	180 & 1.15 & 1 & 9.33 & 16 \\
	180 & 1.15 & 2 & 8.33 & 19 \\
	180 & 1.15 & 5 & 4.00 & 4 \\
	180 & 1.20 & 4 & 12.00 & 12 \\
	180 & 1.20 & 5 & 4.00 & 4 \\
	180 & 1.25 & 2 & 6.50 & 9 \\
	180 & 1.25 & 3 & 2.33 & 3 \\
	180 & 1.25 & 4 & 2.25 & 3 \\
	180 & 1.25 & 5 & 5.25 & 8 \\
	180 & 1.49 & 2 & 6.50 & 10 \\
	180 & 1.49 & 5 & 10.00 & 10 \\
\end{longtable}

\begin{longtable}{P{1cm}P{2cm}P{1cm}P{2cm}P{2.5cm}}
	\caption{Angle between particles, $\alpha$, dimensionless superficial velocity, $U/U_{if}$, label (number) of the test run (set), ensemble-averaged duration for each set, $\left< \Delta t \right>_{set}$, and maximum duration for each set, $\Delta t_{max, set}$, for the clogging regime.} \label{table:tests_Cg}\\
	\toprule
	$\alpha$ ($^{\circ}$) & $U/U_{if}$ & Set & $\left< \Delta t \right>_{set}$ (s)& $\Delta t_{max, set}$ (s)\\ 
	\midrule
	\endfirsthead
	\caption[]{(continued) Angle between particles, $\alpha$, dimensionless superficial velocity, $U/U_{if}$, label (number) of the test run (set), ensemble-averaged duration for each set, $\left< \Delta t \right>_{set}$, and maximum duration for each set, $\Delta t_{max, set}$, for the clogging regime.} \\
	\toprule
	$\alpha$ ($^{\circ}$) & $U/U_{if}$ & Set & $\left< \Delta t \right>_{set}$ (s)& $\Delta t_{max, set}$ (s)\\
	\midrule 
	\endhead
	\midrule \multicolumn{5}{r}{{}} \\
	\endfoot
	\bottomrule
	\endlastfoot
	
	60 & 1.06 & 4 & 49.33 & 74\\
	60 & 1.06 & 5 & 202.00 & 300\\
	120 & 0.94 & 1 & 300.00 & 300\\
	120 & 1.00 & 1 & 300.00 & 300\\
	120 & 1.05 & 1 & 14.00 & 14\\
	120 & 1.05 & 2 & 300.00 & 300\\
	120 & 1.05 & 4 & 51.00 & 51\\
	120 & 1.05 & 5 & 262.00 & 262\\
	120 & 1.11 & 1 & 300.00 & 300\\
	120 & 1.11 & 2 & 116.00 & 116\\
	120 & 1.11 & 3 & 65.50 & 94\\
	120 & 1.11 & 4 & 57.67 & 134\\
	120 & 1.11 & 5 & 16.00 & 16\\
	120 & 1.16 & 1 & 300.00 & 300\\
	120 & 1.16 & 2 & 233.00 & 233\\
	120 & 1.16 & 3 & 29.00 & 32\\
	120 & 1.16 & 4 & 90.00 & 90\\
	120 & 1.16 & 5 & 300.00 & 300\\
	120 & 1.22 & 1 & 300.00 & 300\\
	120 & 1.22 & 2 & 105.00 & 105\\
	120 & 1.22 & 3 & 300.00 & 300\\
	120 & 1.22 & 5 & 137.00 & 137\\
	120 & 1.27 & 5 & 264.00 & 264\\
	120 & 1.33 & 2 & 13.00 & 13\\
	120 & 1.33 & 4 & 14.00 & 14\\
	120 & 1.33 & 5 & 24.00 & 24\\
	150 & 0.94 & 2 & 155.00 & 175\\
	150 & 1.00 & 4 & 270.00 & 270\\
	150 & 1.05 & 4 & 300.00 & 300\\
	150 & 1.05 & 5 & 125.00 & 125\\
	150 & 1.11 & 4 & 146.00 & 285\\
	150 & 1.11 & 5 & 49.50 & 51\\
	150 & 1.16 & 4 & 48.00 & 82\\
	150 & 1.16 & 5 & 39.50 & 49\\
	150 & 1.22 & 4 & 236.00 & 236\\
	150 & 1.22 & 5 & 287.00 & 287\\
	150 & 1.27 & 4 & 283.00 & 283\\
	150 & 1.27 & 5 & 133.00 & 133\\
	150 & 1.33 & 5 & 9.00 & 9\\
	150 & 1.38 & 5 & 30.00 & 34\\
	180 & 0.80 & 4 & 300.00 & 300\\
	180 & 1.00 & 3 & 267.00 & 267\\
	180 & 1.10 & 4 & 247.00 & 247\\
	180 & 1.15 & 5 & 16.00 & 16\\
	180 & 1.25 & 2 & 18.00 & 18\\
\end{longtable}

\begin{longtable}{P{1cm}P{2cm}P{1cm}P{1.5cm}P{1.5cm}P{1.5cm}}
	\caption{ Angle between particles, $\alpha$, dimensionless superficial velocity, $U/U_{if}$, label (number) of the test run (set), initial time of the GT process, $t_0$, final time of the GT process, $t_f$, and duration of the GT process, $\Delta t_{set}$, of each glass transition event observed in each test run. OBS: from $t_f$ on, the GT regime remained until the end of the experiments.} \label{table:tests_GT} \\
	\toprule
	$\alpha$ ($^{\circ}$) & $U/U_{if}$ & Set & $t_0$ (s)& $t_f$ (s)& $\Delta t_{set}$ (s) \\ 
	\midrule
	\endfirsthead
	\caption[]{(continued) Angle between particles, $\alpha$, dimensionless superficial velocity, $U/U_{if}$, label (number) of the test run (set), initial time of the GT process, $t_0$, final time of the GT process, $t_f$, and duration of the GT process, $\Delta t_{set}$, of each glass transition event observed in each test run. OBS: from $t_f$ on, the GT regime remained until the end of the experiments.}\\
	\toprule
	$\alpha$ ($^{\circ}$) & $U/U_{if}$ & Set & $t_0$ (s)& $t_f$ (s)& $\Delta t_{set}$ (s)\\ 
	\midrule
	\endhead
	\midrule \multicolumn{6}{r}{{}} \\
	\endfoot
	\bottomrule
	\endlastfoot
	
	90 & 0.95 & 5 & 11 & 44 & 33\\
	150 & 0.94 & 4 & 0 & 25 & 25\\
	150 & 1.00 & 1 & 20 & 137 & 117\\
	150 & 1.00 & 2 & 20 & 135 & 115\\
	150 & 1.05 & 3 & 89 & 244 & 155\\
	150 & 1.05 & 3 & 40 & 69 & 29\\
	180 & 1.00 & 2 & 19 & 26 & 7\\
	180 & 1.00 & 5 & 7 & 20 & 13\\
	180 & 1.05 & 1 & 18 & 33 & 15\\
	180 & 1.05 & 2 & 17 & 51 & 34\\
	180 & 1.05 & 3 & 70 & 129 & 59\\
	180 & 1.05 & 4 & 9 & 127 & 118\\
	180 & 1.05 & 5 & 46 & 73 & 27\\
	180 & 1.10 & 1 & 150 & 152 & 2\\
	180 & 1.10 & 2 & 153 & 153 & 0\\
	180 & 1.10 & 3 & 5 & 53 & 48\\
	180 & 1.10 & 5 & 131 & 164 & 33\\
	180 & 1.15 & 1 & 35 & 177 & 142\\
	180 & 1.15 & 2 & 103 & 173 & 70\\
	180 & 1.15 & 3 & 8 & 52 & 44\\
	180 & 1.15 & 4 & 93 & 160 & 67\\
	180 & 1.15 & 5 & 130 & 206 & 76\\
	180 & 1.20 & 1 & 145 & 217 & 72\\
	180 & 1.20 & 2 & 144 & 230 & 86\\
	180 & 1.20 & 3 & 50 & 98 & 48\\
	180 & 1.20 & 4 & 224 & 273 & 49\\
	180 & 1.20 & 5 & 197 & 210 & 13\\
	180 & 1.25 & 2 & 219 & 230 & 11\\
\end{longtable}

\begin{table}[h]
	\centering
	\caption{Standard deviations of granular temperatures listed in Tab. \ref{tab:tab2}, for each angle $\alpha$ and superficial velocity $U$.}
	\label{tab:tab2_std}
	\color{black} 
	\begin{tabular}{m{1.5cm} m{1.5cm} m{1.5cm} m{1.5cm} m{1.5cm} m{1.5cm}}
		\hline
		\hline
		$U$ & $\sigma_{\left< \theta_{\alpha,60} \right>}$ & $\sigma_{\left< \theta_{\alpha,90} \right>}$ & $\sigma_{\left< \theta_{\alpha,120} \right>}$ & $\sigma_{\left< \theta_{\alpha,150} \right>}$ & $\sigma_{\left< \theta_{\alpha,180} \right>}$\\
		(cm/s) & (cm/s)$^2$ & (cm/s)$^2$ & (cm/s)$^2$ & (cm/s)$^2$ & (cm/s)$^2$ \\
		\hline
		2.7 & 0.11 & 0.14 & 0.13 & 0.07 & 0.16 \\
		5.5 & 0.12 & 0.05 & 0.17 & 0.08 & 0.04 \\
		6.0 & 0.30 & 0.14 & 0.10 & 0.04 & 0.06 \\
		6.6 & 0.14 & 0.04 & 0.06 & 0.22 & 0.07 \\
		7.1 & 0.14 & 0.12 & 0.06 & 0.11 & 0.06 \\
		7.7 & 0.21 & 0.23 & 0.17 & 0.16 & 0.04 \\
		8.2 & 0.32 & 0.19 & 0.12 & 0.12 & 0.08 \\
		8.8 & 0.09 & 0.10 & 0.15 & 0.39 & 0.08 \\
		9.3 & 0.77 & 0.61 & 0.17 & 0.27 & 0.09 \\
		9.9 & 2.66 & 1.76 & 0.07 & 1.28 & 0.22 \\
		10.4 & 4.98 & 0.10 & 2.86 & 3.48 & 0.24 \\
		11.0 & 6.28 & 3.56 & 0.08 & 4.24 & 0.17 \\
		11.5 & 6.28 & 3.56 & 0.08 & 4.24 & 0.17 \\
		12.1 & 6.46 & 6.63 & 0.09 & 6.32 & 3.21 \\
		12.6 & 6.82 & 6.53 & 6.45 & 5.82 & 2.37 \\
		13.2 & 6.86 & 7.00 & 6.16 & 5.88 & 4.46 \\
		13.7 & 6.75 & 6.51 & 6.00 & 6.00 & 5.02 \\
		16.4 & 6.69 & 6.44 & 5.92 & 6.38 & 5.99 \\
		19.2 & 6.18 & 6.06 & 11.21 & 5.71 & 5.76 \\
		\hline
		\hline
	\end{tabular}
\end{table}

\clearpage

%

\end{document}